\documentclass[aps,prl,twocolumn,groupedaddress,showpacs,floatfix]{revtex4}
\usepackage{graphicx}
\usepackage{psfig}
\usepackage{epsf}
\usepackage{epsfig}
\newcommand{\eq}[1]{Eq.~(\ref{#1})}
\begin{document}

\title{Plasma formation from ultracold Rydberg gases}

\author{T.\ Pohl}
\author{T.\ Pattard}
\author{J.M.\ Rost}

\affiliation{MPI for the Physics of Complex Systems, N{\"o}thnitzer
Str.\ 38, D-01187 Dresden, Germany}

\date{\today}

\begin{abstract}
Recent experiments have demonstrated the spontaneous evolution of a gas of
ultracold Rydberg atoms into an expanding ultracold plasma, as well as the
reverse process of plasma recombination into highly excited atomic states.
Treating the evolution of the plasma on the basis of kinetic equations, while
ionization/excitation and recombination are incorporated using rate equations,
we have investigated theoretically the Rydberg-to-plasma transition.
Including the
influence of spatial correlations on the plasma dynamics in an approximate way 
we find that ionic correlations change the results only quantitatively but not
qualitatively.
\end{abstract}


\maketitle

Advances in cooling and trapping of neutral gases have opened up a new branch of
atomic physics, namely dynamics in ultracold ($T \ll 1$ K) systems.
One interesting topic is the physics of ultracold Rydberg gases and ultracold
neutral plasmas. In  recent experiments, ultracold neutral plasmas have
been produced from a small cloud of laser-cooled atoms confined in a
magneto-optical trap \cite{Kil99,Kul00,Kil01,Rob00,Eyl00}. In the experiments
performed at NIST \cite{Kil99,Kul00,Kil01}, a plasma was produced by
photoionizing laser-cooled Xe atoms with an initial ion temperature of
about $10
\, \mu {\rm K}$. By tuning the frequency of the ionizing laser, the initial
electron energy $E_e$ was varied corresponding to a temperature
range $1 \, {\rm
K} < E_e / k_\mathrm{B} < 1000 \, {\rm K}$, and the subsequent expansion of the
plasma into the surrounding vacuum was studied systematically. Remarkably,  a
significant amount of recombination was observed as the plasma expands, leading
to the formation of Rydberg atoms from the plasma. In a complementary type of
experiment \cite{Rob00,Eyl00}, ultracold atoms were laser-excited into high
Rydberg states rather than directly ionized. In these experiments, also the
reverse process has been observed, namely the spontaneous evolution of the
Rydberg gas into a plasma.

An issue in the theoretical considerations \cite{Kuz02,Rob02,Maz02,Rus}
stimulated by the experiments has been the question whether the evolving plasma
is strongly coupled or not.
Due to the very low initial temperature, the electron
Coulomb coupling parameter $\Gamma_e (t=0)$ is found to be significantly larger
than one ($\Gamma_e = e^2/(a k_\mathrm{B} T_e)$, where $a$ is the Wigner-Seitz
radius).
On the other hand, it has been pointed out in \cite{Kuz02} that, for the
initial conditions of the NIST experiments \cite{Kil99,Kul00,Kil01}, the
development of equilibrium electron-electron correlations leads to a rapid
heating of the electron gas, which brings $\Gamma_e$ down to order unity on a
very short timescale. In \cite{Kuz02}, this was demonstrated by a
molecular-dynamics simulation of the electron and ion motion. The 
calculation was limited
to a short time interval ($\approx$ 100 nanoseconds)
in the initial stage of the plasma
evolution due to the large numerical effort required.
For the first quantitative comparison with experiment 
the plasma dynamics has been modeled within a kinetic approach,
while ionization,
excitation and recombination has been treated by a separate set of rate
equations \cite{Rob02}. The electron dynamics was described in an adiabatic 
approximation
with the atom and ion temperatures set to zero. Since this model does
only account for the mean-field potential created by the charges, possible
correlation heating could not be described. Nevertheless,
 due to heating by three-body recombination events,
$\Gamma_e$ does not exceed a value of $\approx 0.2$ during the plasma expansion.
Hence, the influence of electronic correlations on the plasma dynamics
could be neglected.  With zero ionic temperature the {\em ionic}
Coulomb coupling parameter, however, is infinite in the framework of
this model.  Consequently, the role of ion-ion correlations could not
be estimated.

Our description is similar to that of \cite{Rob02} with small differences, e.g.,
inclusion of black-body radiation as a source for photoionization. This
is necessary to describe the initial ionization of the Rydberg gas in the
experiments \cite{Rob00,Eyl00} since the ionization rate through cold atom-atom
collisions is much lower at these densities.
More importantly, in addition to \cite{Rob02}  we allow for
possible spatial correlation effects. Briefly, we use a set of kinetic equations
for the plasma expansion, combined with rate equations for the description of
ionization/excitation and recombination. The kinetic equations are derived from
the first equation of the BBGKY hierarchy, which yields the standard Vlasov
equation for the evolution of a collisionless plasma, augmented by additional
terms accounting for spatial correlations. Since the electronic Coulomb coupling
parameter is well below unity under all experimental conditions realized so far
(as
we have convinced ourselves including electronic correlations on the basis of
two-component Debye-H\"uckel theory; see also \cite{Rob02}) we
neglect them and treat the ions as a one-component plasma embedded in a
neutralizing background. 

Assuming that the correlation function vanishes at
distances larger than the correlation length $a_c$ and that the spatial density
varies slowly over a distance $a_c$, we find that the force induced by ionic
correlations can be  expressed in terms of the local correlation energy per
particle
\begin{equation}
\label{ecorr}
u_{ii}({\mathbf{r}})=\frac{e^{2}}{2}\rho_{i}({\mathbf{r}})\int
d{\mathbf{y}}\frac{g_{ii}(y ; {\mathbf{r}})}{y} \;\;\; ,
\end{equation}
where the spatial density $\rho_i$ is related to the one-particle distribution
function $f_i$ by \(\rho_{i}=\int d{\mathbf{v}}f_{i}\) and $g(y)$ is the
correlation function.  The correlation energy for a one-component
plasma in thermal equilibrium is a function of the Coulomb coupling
parameter only \cite{Dub99}, and simple analytical formulae have been
given for a wide range of $\Gamma$.  We use 
\begin{equation} \label{ucorr}
u_{ii}^{\mathrm{(eq)}}=-\frac{9}{10}k_{\mathrm{B}}T_i
\Gamma_{i}=-
\frac{9 e^2}{10}\left(\frac{4}{3}\pi\rho_{i}\right)^{1/3} \;\;\; ,
\end{equation}
derived
from the ion sphere model
which is a good approximation for $\Gamma > 1$ \cite{Ich82}.

The timescale for relaxation towards the individual equilibrium of
electrons and ions, respectively, is determined by the corresponding
inverse plasma frequency
\(\omega_{{\mathrm{p}}}^{-1}=\sqrt{m_{}/4\pi
\rho_{}e^{2}}\) \cite{Mur01}.  Under typical conditions of the
experiments \cite{Kil99,Kul00,Rob00,Kil01}, it
is of the order of $10^{-9}$s for electrons, while equilibration
of the ions takes between $10^{-7}$s and $10^{-5}$s depending on $\rho$.
These timescales for equilibration have to be seen relative to the expansion
time of the plasma, which is of the order of
several $10^{-6}$s. Hence, an adiabatic
approximation for the electron distribution function can be safely applied,
such that the kinetic
equation for the electronic component of the plasma yields a relation
allowing one to express the mean-field potential in terms of the
density $\rho_e$, which can be inserted into the corresponding
equation for the ions.

The much slower equilibration of the ions 
renders an adiabatic approximation {\em a priori} difficult. However, even in
situations where static parameters, e.g., relevant masses, periods or 
rates, speak against an adiabatic
treatment it is sometimes made possible through dynamical adiabaticity
(see, e.g., \cite{TRR00}). Of course, this can only be verified {\em a
posteriori} by more elaborate calculations without
adiabatic approximation.  For now, we simply assume an adiabatic time
evolution of the plasma, i.e.\ we estimate the ionic correlation energy
from its equilibrium value eq.\ (\ref{ucorr}) and use a homogeneous ion
temperature at all times.  This is the
opposite limit to a zero ionic temperature which would never lead to
equilibration.  Hence, our results are expected to clarify whether
(and to what extent) correlation effects can have any influence on the
dynamics at all.

The large difference in timescales between ionic and electronic motion 
justifies the quasineutral approximation $\rho_e=\rho_i=\rho$
\cite{Dor98} applicable under the
experimental conditions in \cite{Kil99,Kul00,Rob00,Kil01}. Under these 
circumstances  a closed
equation for the ion distribution function is obtained and the resulting
equation {\it without the correlation term} permits selfsimilar
analytical solutions \cite{Dor98}.  One of them is the gaussian
profile describing the initial state in the experiments under
consideration.  Exact selfsimilar solutions
exist even in the case of an additional linear force \cite{Dor00}. 
This condition is satisfied by the correlation force to a good
approximation save for the outer periphery of the plasma \cite{PPR03}. 
Hence, we can write
\begin{equation}
\label{ansatz}
f_i \propto \exp\left(-\frac{r^2}{2\sigma^2}
-\frac{m_i\left({\mathbf{v}}-{\mathbf{w}}({\mathbf{r}})\right)^2}
{2k_{\mathrm{B}}T_i}\right) \;\;\; ,
\end{equation}
where $\sigma$ is the width of the spatial density distribution and
$\mathbf{w}({\mathbf{r}}) = \gamma \mathbf{r}$ is the hydrodynamic velocity
(see also\cite{Rob02}). 
Substitution of \eq{ansatz} finally leads to the set of equations
\begin{eqnarray}
\label{allevolutions}
\frac{d\sigma^2}{dt}&=&2\gamma \sigma^2
 \\
\frac{d\gamma}{dt}+\gamma^2&=&\frac{N_e}{M\sigma^2}\left[k_{\mathrm{B}}
\left(T_e+T_i\right)+W_c\right], \nonumber
\end{eqnarray}
where
\(W_{c}=1/3\int d{\mathbf{r}} \rho \partial (u_{ii}\rho)/\partial\rho\)
arises from the correlation pressure and $M$ is the total mass of the
plasma.  The thermal energy is determined from the total energy of the
system
\begin{equation} \label{encons}
E_{\mathrm{tot}} = \frac{3}{2}N_{{e}}k_{\mathrm{B}}\left(T_{{e}}+
T_{{i}}\right)+\frac{3}{2}M\gamma^2\sigma^2+ U_c \;\; ,
\end{equation}
i.e.\ the sum of the kinetic energy and the correlation energy \(U_c=\int
d{\mathbf{r}} u_{ii}\rho\).
The set of equations (\ref{allevolutions}) reduces to the one used in
\cite{Rob02} if the correlation term is dropped and $T_i\equiv 0$.  
More recently, it has been shown
 by comparison with  more
sophisticated calculations that this simple ansatz provides a good
description of the plasma dynamics even if some of the assumptions,
such as the quasineutrality of the plasma and the gaussian
distribution of the densities, only hold approximately \cite{Rob03} .

The low-temperature enhancement of the expansion
velocity of the ultracold plasma produced in \cite{Kul00} can be well described
by combining a hydrodynamic description of the
expansion dynamics with conventional rate equations
accounting for inelastic collisions between the plasma particles and
Rydberg atoms \cite{Rob02}. As a result,
the total kinetic energy becomes a function of time since the Rydberg atoms
act as energy sinks and sources.
In our calculations, we include bound-bound transitions by electron impact
excitation and deexcitation, and bound-free transitions by three-body
recombination, electron impact ionization, and black-body radiation.
We use the collision rate coefficients
derived by Mansbach and Keck \cite{Man69} and
black-body photoionization is described by
first-order perturbation theory \cite{Gal94}, using an
asymptotic expression for the atomic oscillator strengths \cite{Joh72}.

Finally, the question of the expansion velocity of the Rydberg atoms
has to be addressed, since they are not driven by the Coulomb drag of
the expanding electrons.  Under typical experimental conditions
(\(\rho_{{e}}=10^9 {\rm cm}^{-3}\), \(T_{{e}}=20\) K), the timescale
for electron-atom-collisions is of the order of $10^{-8}$s.
Therefore, in the course of the expansion of the plasma ($10^{-5}$s)
any given atom
with binding energy of the order of several $k_\mathrm{B} T$ will
recombine and re-ionize many times, changing its character between ion
and Rydberg atom.  In addition, collisions between ions and atoms
significantly equilibrate the hydrodynamical velocities even for
lower Rydberg states \cite{Nem02}.  Following this reasoning we
assume equal hydrodynamical velocities and density profiles for the
ions and atoms. This implies that the expansion of the neutral Rydberg
atoms can be simply taken into account by 
changing the  mass of the ions  to an effective mass  $M$ in 
\eq{allevolutions}  which is the
total mass of the {\em total} system (plasma + atoms).

We have simulated the expansion of an initially fully ionized plasma as
reported in \cite{Kul00,Kil01}. In accordance with previous calculations
\cite{Rob02}, we find quantitative agreement for the electron energy dependence
of the asymptotic expansion velocity. Furthermore, the calculations
qualitatively reproduce the nonmonotonic time dependence of the number
of {\em detected}
atoms observed in \cite{Kil01}. We  
note that the ionic correlations do not significantly
change the {\em plasma} dynamics for these experiments.

The results of our calculations for the experiments \cite{Rob00,Eyl00} are
summarized in figures 1 to 3.
\begin{figure}
\epsfysize 4.5cm
\centerline{\epsffile{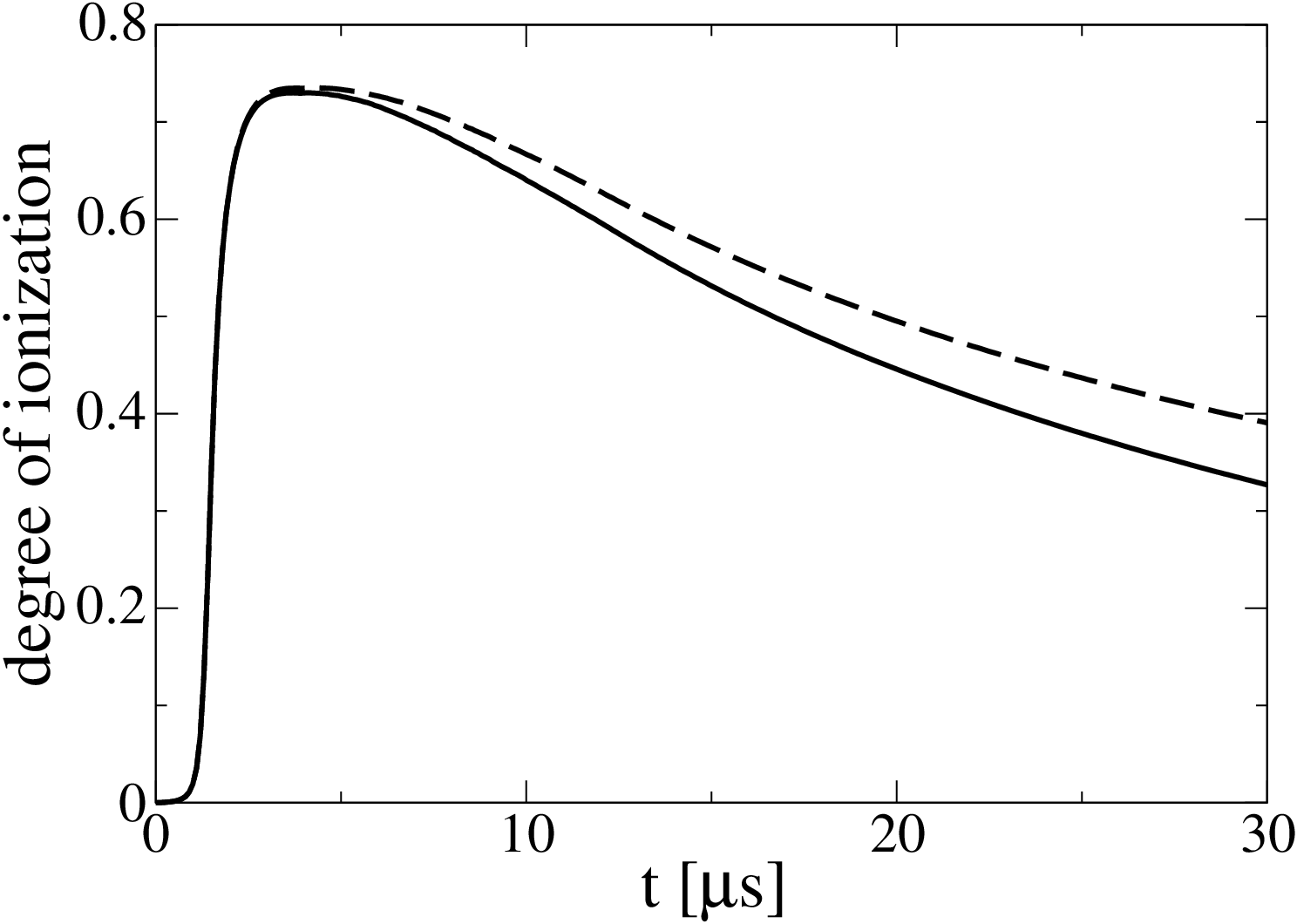}}
\caption{\label{fig1} Time evolution of the degree of ionization for
the following initial conditions: density of atoms
$\rho=8\cdot10^{9}\:{\rm cm}^{-3}$, atomic temperature $T_a=140\:\mu
{\rm K}$, plasma width \(\sigma=60\:\mu {\rm m}\) and initial
principal quantum number of the Rydberg atoms \(n_0=70\).  Results are
shown with ionic correlations (solid) and without (dashed).}
\end{figure}
Figure 1 shows the degree of ionization of the Rydberg gas, i.e.\ the
plasma fraction of the system, as a function of time during the
expansion of the system. For comparison, the result without correlation
($U_c = T_i \equiv 0$ in eqs.\ (\ref{allevolutions}) and (\ref{encons}))
is also shown.
As has been observed experimentally, it
takes of the order of two to three microseconds before significant
plasma formation occurs which sets in with an ionization avalanche.
Note that the early time development including the ionization 
avalanche is obviously not influenced by the ionic correlations. 
Their development  requires
a significant amount of charged plasma particles  which are only 
available after the ionization avalanche. The production of initial
``seed charges''   by black-body radiation 
does not depend on ionic correlations. 
The ionization avalanche appears somewhat ($\approx 1 \mu$s) earlier
than in the experiment since in reality the first electrons
produced by the black-body radiation will leave the atom cloud until a
sufficiently strong positive space charge builds up to trap the electrons.
With the increasing number of free charges,
the plasma forms quickly, followed by a partial back evolution
into a Rydberg gas. Fig.~\ref{fig1} shows that this part of the gas
dynamics may indeed be affected by  ion-ion correlation, which
leads to a more efficient recombination in the final stage of the
plasma evolution. The reason for this behavior can be understood from
eqs.\ (\ref{allevolutions}) and (\ref{encons}). Together,
$W_c$ and $U_c$ lead to an additional acceleration
($2/9 \,U_c$ if \eq{ucorr} is used for $u$) in addition to
the ideal thermal electron pressure \(k_\mathrm{B} T_e\). Hence, the
adiabatic electron cooling during the expansion is faster than without
correlations. In turn recombination is accelerated, since the corresponding rate
strongly increases with decreasing temperature.
We note that corresponding effects of the same order of magnitude are also
found for the experiments \cite{Kul00,Kil01}. The reason
that they do not significantly affect the expansion dynamics of the plasma part
lies in the fact that the correlations mainly influence the recombination
into high-lying states at later stages of the evolution. Since these are
very weakly bound, they hardly influence the kinetic energy of the system
which determines the expansion velocity.
Hence, good agreement is found between calculations
with and without inclusion of correlation effects as long as quantities
related to the plasma expansion are compared.

\begin{figure}
\centerline{\epsfysize = 5.5cm \epsffile{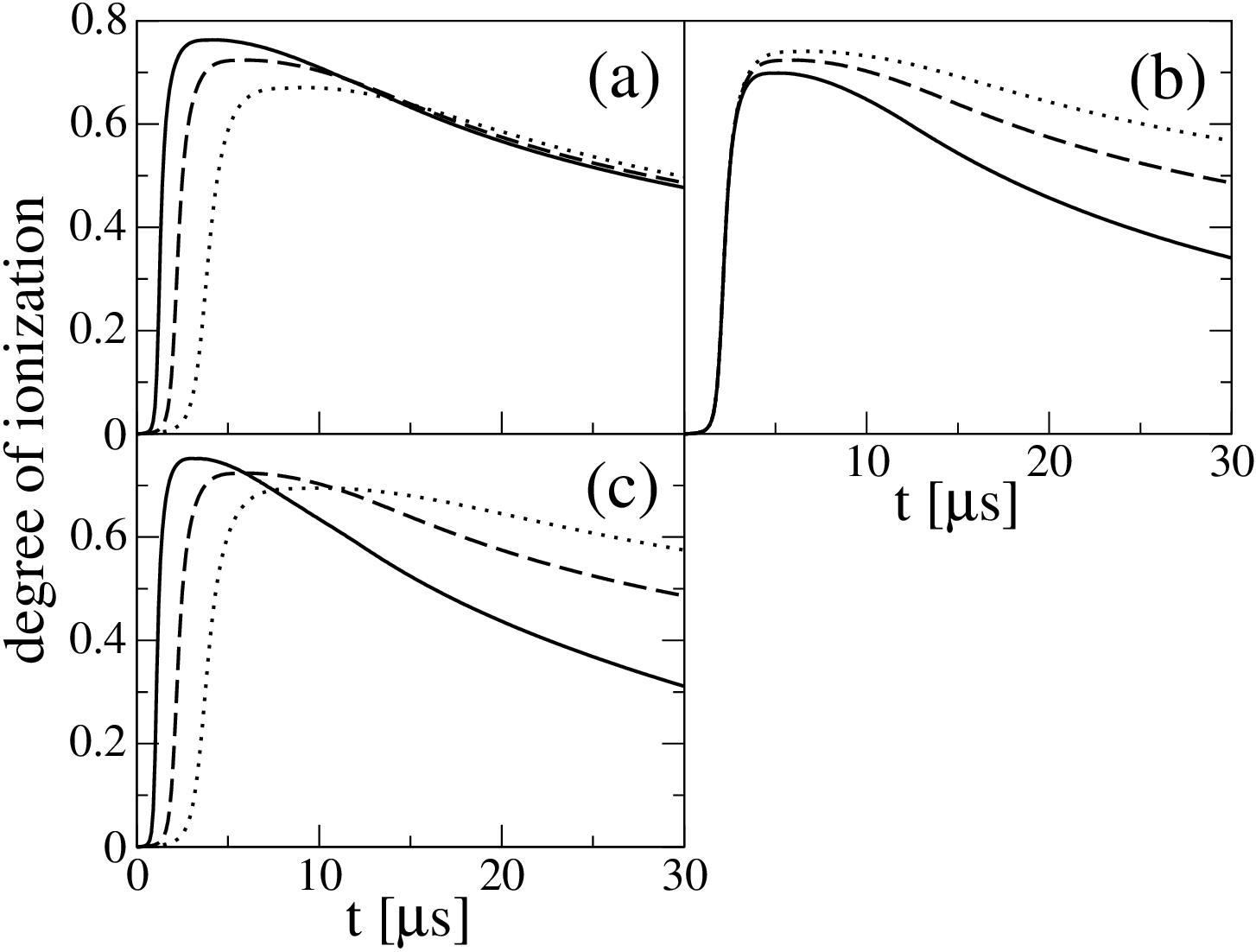}}
\caption{\label{fig2} 
Degree of ionization as a function of time for various initial conditions
($n_0 = 70$).
a) $\sigma=80\:\mu {\rm m}$ fixed, $\rho= 2.5 \cdot10^{9}$ (dotted),
$5\cdot10^{9}$ (dashed), $1\cdot10^{10}$ cm$^{-3}$ (solid);
b) $\rho= 5\cdot10^{9}$ cm$^{-3}$ fixed, $\sigma=60$ (solid), $80$ (dashed),
1$00\:\mu {\rm m}$ (dotted);
c) $N_a=114000$ fixed, $\sigma=60$ (solid), $80$ (dashed), $100\:\mu {\rm m}$
(dotted).}
\end{figure}
Figure \ref{fig2}
illustrates the dependence of the plasma evolution on the initial
conditions. For fixed initial width of the Rydberg atom cloud 
(Fig.~\ref{fig2}a),
the plasma formation occurs earlier with increasing density, due to the
fact that ionization is faster and more efficient at higher densities. The
later stages of the evolution, on the other hand, depend only weakly
on the density. This may be attributed to the fact that in the evolution
equation (\ref{allevolutions}), $\rho$ enters only in the form $N/M$, i.e.\
essentially as the degree of ionization.
For fixed initial density $\rho$,
the ionization avalanche occurs at the same time, in accordance with the
argument made previously (Fig.~\ref{fig2}b). However, for smaller width,
the plasma expands faster,
leading ultimately to lower electron temperatures and increased recombination.
Finally, if the
number of atoms $N_{a}$ is kept fixed the density changes when the 
size of the atom cloud is varied. Hence, Fig.~\ref{fig2}c can be 
interpreted as the combined effect seen in Figs.~\ref{fig2}a,b: The 
ionization avalanche occurs at different times for each curve (due to 
different atomic densities) and the relaxation leads to different 
final degrees of ionization (due to the different size of the atomic 
cloud). 

\begin{figure}
\centerline{\epsfysize = 6cm \epsffile{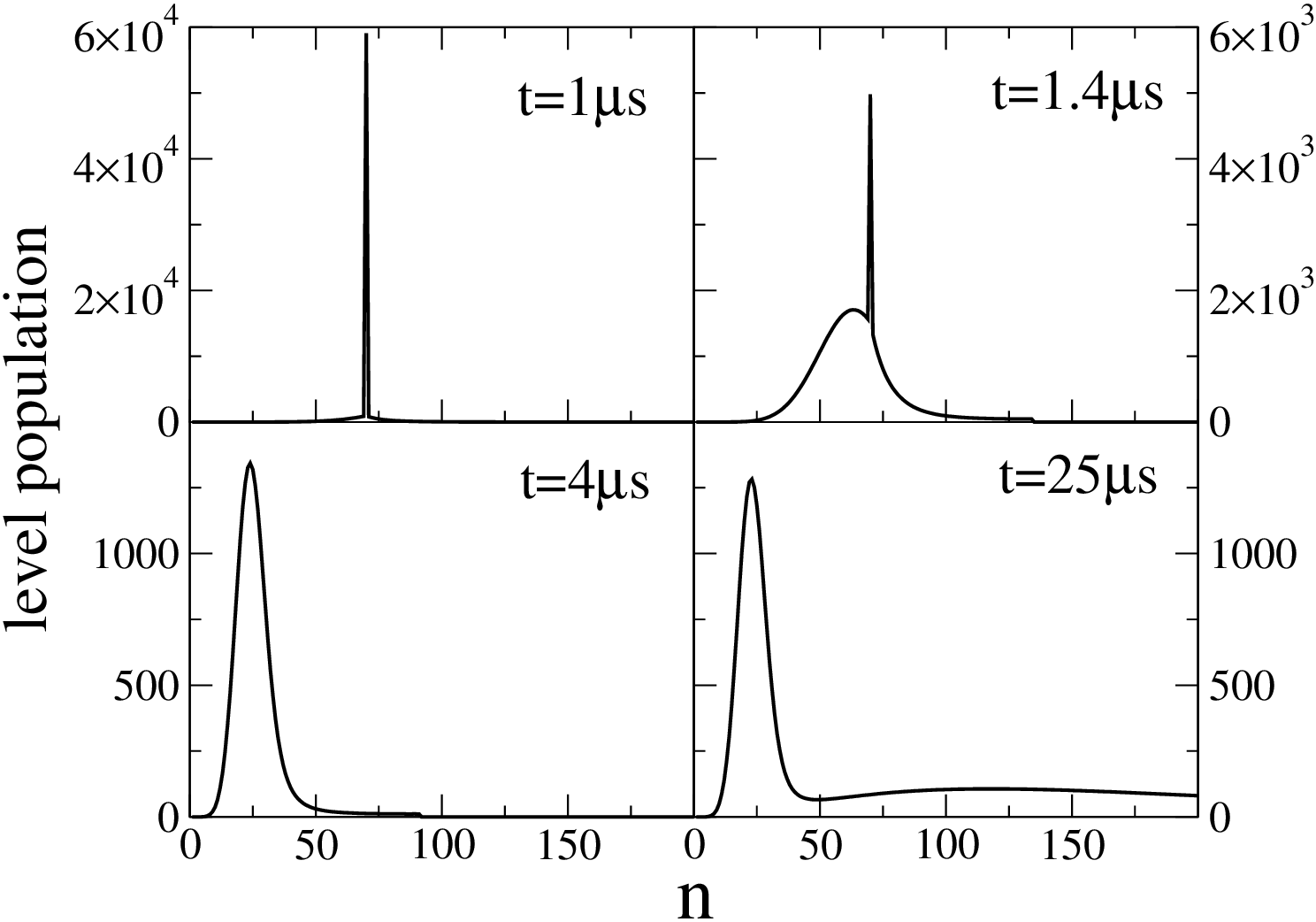}}
\caption{\label{fig3} Level distribution of Rydberg atoms after \(1\mu {\rm
s}\), \(1.4\mu {\rm s}\), \(4 \mu {\rm s}\) and \(25 \mu {\rm s}\).}
\end{figure}
Figure \ref{fig3} shows the non-ionized part of the gas, more precisely the
population of Rydberg levels in the initial and final stage of the expansion,
as well as at the time of the maximum degree of ionization. Initially, all atoms
are prepared with a principal quantum number $n_0 = 70$. At later times,
the interplay between ionization, recombination and exciting and deexciting
collisions leads to a decrease of the average excitation.
This decrease in excitation is the main source of energy
that triggers the ionization and the plasma expansion, energy absorbed from
the black-body radiation field is negligible in comparison. Finally,
recombination repopulates higher levels (which ``freeze out'' due to the
rapid decrease of electron temperature), but the peak of the distribution is
still found at relatively low excitations.
It would be very interesting to measure the final distribution of Rydberg
atoms, which should provide a stringent test for theoretical
models.

In summary, we have described the formation of a cold plasma
from a highly excited Rydberg gas, using a simple model based on kinetic
equations for the plasma evolution. Microscopic molecular-dynamics
simulations are limited to much shorter timescales due to the large numerical
effort required. We have estimated
the maximum effect of ionic correlations by including them into the model
assuming instant equilibration of the ions. In this respect the calculations
without equilibration (no correlation, ionic temperature zero) and with 
instant equilibration should bracket the actual ionization yield of the Rydberg
gas. Given the difference in the
ionization yields with and without ionic correlation in fig.\ \ref{fig1},
more elaborate
calculations are worthwhile although at present it is unclear
whether the plasma evolution including correlation can be followed
over sufficently long times,
e.g., with P$^3$M-codes.

\begin{acknowledgments}
It is a pleasure to thank P.\ Gould, E.\ Eyler, T.\ Killian and S.\ Kulin
for helpful
comments. Financial support by DFG through SPP 1116 is also gratefully
acknowledged.
\end{acknowledgments}


\begin{thebibliography}{21}
\expandafter\ifx\csname natexlab\endcsname\relax\def\natexlab#1{#1}\fi
\expandafter\ifx\csname bibnamefont\endcsname\relax
  \def\bibnamefont#1{#1}\fi
\expandafter\ifx\csname bibfnamefont\endcsname\relax
  \def\bibfnamefont#1{#1}\fi
\expandafter\ifx\csname citenamefont\endcsname\relax
  \def\citenamefont#1{#1}\fi
\expandafter\ifx\csname url\endcsname\relax
  \def\url#1{\texttt{#1}}\fi
\expandafter\ifx\csname urlprefix\endcsname\relax\def\urlprefix{URL }\fi
\providecommand{\bibinfo}[2]{#2}
\providecommand{\eprint}[2][]{\url{#2}}

\bibitem[{\citenamefont{Killian et~al.}(1999)\citenamefont{Killian, Kulin,
  Bergeson, Orozco, Orzel, and Rolston}}]{Kil99}
\bibinfo{author}{\bibfnamefont{T.C.}~\bibnamefont{Killian}},
  \bibinfo{author}{\bibfnamefont{S.}~\bibnamefont{Kulin}},
  \bibinfo{author}{\bibfnamefont{S.D.}~\bibnamefont{Bergeson}},
  \bibinfo{author}{\bibfnamefont{L.A.}~\bibnamefont{Orozco}},
  \bibinfo{author}{\bibfnamefont{C.}~\bibnamefont{Orzel}}, \bibnamefont{and}
  \bibinfo{author}{\bibfnamefont{S.L.}~\bibnamefont{Rolston}},
  \bibinfo{journal}{Phys.\ Rev.\ Lett.} \textbf{\bibinfo{volume}{83}},
  \bibinfo{pages}{4776} (\bibinfo{year}{1999}).

\bibitem[{\citenamefont{Kulin et~al.}(2000)\citenamefont{Kulin, Killian,
  Bergeson, and Rolston}}]{Kul00}
\bibinfo{author}{\bibfnamefont{S.}~\bibnamefont{Kulin}},
  \bibinfo{author}{\bibfnamefont{T.C.}~\bibnamefont{Killian}},
  \bibinfo{author}{\bibfnamefont{S.D.}~\bibnamefont{Bergeson}},
  \bibnamefont{and}
  \bibinfo{author}{\bibfnamefont{S.L.}~\bibnamefont{Rolston}},
  \bibinfo{journal}{Phys.\ Rev.\ Lett.} \textbf{\bibinfo{volume}{85}},
  \bibinfo{pages}{318} (\bibinfo{year}{2000}).

\bibitem[{\citenamefont{Killian et~al.}(2001)\citenamefont{Killian, Lim, Kulin,
  Dumke, Bergeson, and Rolston}}]{Kil01}
\bibinfo{author}{\bibfnamefont{T.C.}~\bibnamefont{Killian}},
  \bibinfo{author}{\bibfnamefont{M.J.}~\bibnamefont{Lim}},
  \bibinfo{author}{\bibfnamefont{S.}~\bibnamefont{Kulin}},
  \bibinfo{author}{\bibfnamefont{R.}~\bibnamefont{Dumke}},
  \bibinfo{author}{\bibfnamefont{S.D.}~\bibnamefont{Bergeson}},
  \bibnamefont{and}
  \bibinfo{author}{\bibfnamefont{S.L.}~\bibnamefont{Rolston}},
  \bibinfo{journal}{Phys.\ Rev.\ Lett.} \textbf{\bibinfo{volume}{86}},
  \bibinfo{pages}{3759} (\bibinfo{year}{2001}).

\bibitem[{\citenamefont{Robinson et~al.}(2000)\citenamefont{Robinson, Tolra,
  Noel, Gallagher, and Pillet}}]{Rob00}
\bibinfo{author}{\bibfnamefont{M.P.}~\bibnamefont{Robinson}},
  \bibinfo{author}{\bibfnamefont{B.L.}~\bibnamefont{Tolra}},
  \bibinfo{author}{\bibfnamefont{M.W.}~\bibnamefont{Noel}},
  \bibinfo{author}{\bibfnamefont{T.F.}~\bibnamefont{Gallagher}},
  \bibnamefont{and} \bibinfo{author}{\bibfnamefont{P.}~\bibnamefont{Pillet}},
  \bibinfo{journal}{Phys.\ Rev.\ Lett.} \textbf{\bibinfo{volume}{85}},
  \bibinfo{pages}{4466} (\bibinfo{year}{2000}).

\bibitem[{\citenamefont{Eyler et~al.}(2000)\citenamefont{Eyler, Estrin, Ensher,
  Cheng, Sanborn, and Gould}}]{Eyl00}
\bibinfo{author}{\bibfnamefont{E.}~\bibnamefont{Eyler}},
  \bibinfo{author}{\bibfnamefont{A.}~\bibnamefont{Estrin}},
  \bibinfo{author}{\bibfnamefont{J.R.}~\bibnamefont{Ensher}},
  \bibinfo{author}{\bibfnamefont{C.H.}~\bibnamefont{Cheng}},
  \bibinfo{author}{\bibfnamefont{C.}~\bibnamefont{Sanborn}}, \bibnamefont{and}
  \bibinfo{author}{\bibfnamefont{P.L.}~\bibnamefont{Gould}},
  \bibinfo{journal}{Bull.\ Am.\ Phys.\ Soc.} \textbf{\bibinfo{volume}{45}},
  \bibinfo{pages}{56} (\bibinfo{year}{2000}).

\bibitem[{\citenamefont{Kuzmin and O'Neil}(2002)}]{Kuz02}
\bibinfo{author}{\bibfnamefont{S.G.}~\bibnamefont{Kuzmin}} \bibnamefont{and}
  \bibinfo{author}{\bibfnamefont{T.M.}~\bibnamefont{O'Neil}},
  \bibinfo{journal}{Phys.\ Rev.\ Lett.} \textbf{\bibinfo{volume}{88}},
  \bibinfo{pages}{065003} (\bibinfo{year}{2002}).

\bibitem[{\citenamefont{Robicheaux and Hanson}(2002)}]{Rob02}
\bibinfo{author}{\bibfnamefont{F.}~\bibnamefont{Robicheaux}} \bibnamefont{and}
  \bibinfo{author}{\bibfnamefont{J.D.}~\bibnamefont{Hanson}},
  \bibinfo{journal}{Phys.\ Rev.\ Lett.} \textbf{\bibinfo{volume}{88}},
  \bibinfo{pages}{055002} (\bibinfo{year}{2002}).

\bibitem[{\citenamefont{Mazevet et~al.}(2002)\citenamefont{Mazevet, Collins,
  and Kress}}]{Maz02}
\bibinfo{author}{\bibfnamefont{S.}~\bibnamefont{Mazevet}},
  \bibinfo{author}{\bibfnamefont{L.A.}~\bibnamefont{Collins}}, \bibnamefont{and}
  \bibinfo{author}{\bibfnamefont{J.D.}~\bibnamefont{Kress}},
  \bibinfo{journal}{Phys.\ Rev.\ Lett.} \textbf{\bibinfo{volume}{88}},
  \bibinfo{pages}{055001} (\bibinfo{year}{2002}).

\bibitem[{\citenamefont{Tkachev and Yakovlenko}(2001)}]{Rus}
\bibinfo{author}{\bibfnamefont{A.N.}~\bibnamefont{Tkachev}} \bibnamefont{and}
  \bibinfo{author}{\bibfnamefont{S.I.}~\bibnamefont{Yakovlenko}},
  \bibinfo{journal}{Quantum Electronics} \textbf{\bibinfo{volume}{31}},
  \bibinfo{pages}{1084} (\bibinfo{year}{2001}).

\bibitem[{\citenamefont{Dubin and O'Neil}(1999)}]{Dub99}
\bibinfo{author}{\bibfnamefont{D.H.E.}~\bibnamefont{Dubin}} \bibnamefont{and}
  \bibinfo{author}{\bibfnamefont{T.M.}~\bibnamefont{O'Neil}},
  \bibinfo{journal}{Rev.\ Mod.\ Phys.} \textbf{\bibinfo{volume}{71}},
  \bibinfo{pages}{87} (\bibinfo{year}{1999}).

\bibitem[{\citenamefont{Ichimaru}(1982)}]{Ich82}
\bibinfo{author}{\bibfnamefont{S.}~\bibnamefont{Ichimaru}},
  \bibinfo{journal}{Rev.\ Mod.\ Phys.} \textbf{\bibinfo{volume}{54}},
  \bibinfo{pages}{1017} (\bibinfo{year}{1982}).

\bibitem[{\citenamefont{Murillo}(2001)}]{Mur01}
\bibinfo{author}{\bibfnamefont{M.S.}~\bibnamefont{Murillo}},
  \bibinfo{journal}{Phys.\ Rev.\ Lett.} \textbf{\bibinfo{volume}{87}},
  \bibinfo{pages}{115003} (\bibinfo{year}{2001}).

\bibitem[{\citenamefont{Tanner et~al.}(2000)\citenamefont{Tanner, Richter, and
  Rost}}]{TRR00}
\bibinfo{author}{\bibfnamefont{G.}~\bibnamefont{Tanner}},
  \bibinfo{author}{\bibfnamefont{K.}~\bibnamefont{Richter}}, \bibnamefont{and}
  \bibinfo{author}{\bibfnamefont{J.M.}~\bibnamefont{Rost}},
  \bibinfo{journal}{Rev.\ Mod.\ Phys.} \textbf{\bibinfo{volume}{72}},
  \bibinfo{pages}{497} (\bibinfo{year}{2000}).

\bibitem[{\citenamefont{Dorozhkina and Semenov}(1998)}]{Dor98}
\bibinfo{author}{\bibfnamefont{D.S.}~\bibnamefont{Dorozhkina}} \bibnamefont{and}
  \bibinfo{author}{\bibfnamefont{V.E.}~\bibnamefont{Semenov}},
  \bibinfo{journal}{Phys.\ Rev.\ Lett.} \textbf{\bibinfo{volume}{81}},
  \bibinfo{pages}{2691} (\bibinfo{year}{1998}).

\bibitem[{\citenamefont{Dorozhkina and Semenov}(1999)}]{Dor00}
\bibinfo{author}{\bibfnamefont{D.S.}~\bibnamefont{Dorozhkina}} \bibnamefont{and}
  \bibinfo{author}{\bibfnamefont{V.E.}~\bibnamefont{Semenov}},
  \bibinfo{journal}{J. Exp.\ Theor.\ Phys.} \textbf{\bibinfo{volume}{89}},
  \bibinfo{pages}{468} (\bibinfo{year}{1999}).

\bibitem[{\citenamefont{Pohl et~al.}()\citenamefont{Pohl, Pattard, and
  Rost}}]{PPR03}
\bibinfo{author}{\bibfnamefont{T.}~\bibnamefont{Pohl}},
  \bibinfo{author}{\bibfnamefont{T.}~\bibnamefont{Pattard}}, \bibnamefont{and}
  \bibinfo{author}{\bibfnamefont{J.M.}~\bibnamefont{Rost}},
  \bibinfo{howpublished}{to be published}.

\bibitem[{\citenamefont{Robicheaux and Hanson}(2003)}]{Rob03}
\bibinfo{author}{\bibfnamefont{F.}~\bibnamefont{Robicheaux}} \bibnamefont{and}
  \bibinfo{author}{\bibfnamefont{J.D.}~\bibnamefont{Hanson}},
  \bibinfo{howpublished}{preprint} (\bibinfo{year}{2003}).

\bibitem[{\citenamefont{Mansbach and Keck}(1969)}]{Man69}
\bibinfo{author}{\bibfnamefont{P.}~\bibnamefont{Mansbach}} \bibnamefont{and}
  \bibinfo{author}{\bibfnamefont{J.}~\bibnamefont{Keck}},
  \bibinfo{journal}{Phys.\ Rev.} \textbf{\bibinfo{volume}{181}},
  \bibinfo{pages}{275} (\bibinfo{year}{1969}).

\bibitem[{\citenamefont{Gallagher}(1994)}]{Gal94}
\bibinfo{author}{\bibfnamefont{T.F.}~\bibnamefont{Gallagher}},
  \emph{\bibinfo{title}{Rydberg Atoms}} (\bibinfo{publisher}{Cambridge
  University Press}, \bibinfo{year}{1994}).

\bibitem[{\citenamefont{Johnson}(1972)}]{Joh72}
\bibinfo{author}{\bibfnamefont{L.C.}~\bibnamefont{Johnson}},
  \bibinfo{journal}{Astrophys.\ J.} \textbf{\bibinfo{volume}{174}},
  \bibinfo{pages}{227} (\bibinfo{year}{1972}).

\bibitem[{\citenamefont{Nemirovsky et~al.}(2002)\citenamefont{Nemirovsky,
  Fredkin, and Ron}}]{Nem02}
\bibinfo{author}{\bibfnamefont{R.A.}~\bibnamefont{Nemirovsky}},
  \bibinfo{author}{\bibfnamefont{D.R.}~\bibnamefont{Fredkin}}, \bibnamefont{and}
  \bibinfo{author}{\bibfnamefont{A.}~\bibnamefont{Ron}},
  \bibinfo{journal}{Phys.\ Rev.\ E} \textbf{\bibinfo{volume}{66}},
  \bibinfo{pages}{066405} (\bibinfo{year}{2002}).

\end{thebibliography}
\end{document}